\documentclass[twocolumn,aps,amssymb,noshowpacs,superscriptaddress,nofootinbib,longbibliography]{revtex4-1}
\usepackage{amsmath,amssymb}
\usepackage{graphicx}
\usepackage{verbatim}
\usepackage{amsfonts}
\usepackage{dcolumn}
\usepackage{bm}
\usepackage[section]{placeins}
\usepackage{color}
\usepackage{mathrsfs}
\usepackage[colorlinks,bookmarks=true,citecolor=blue,linkcolor=red,urlcolor=blue]{hyperref}
\usepackage{afterpage}
\usepackage{ulem}
\usepackage{url} 
\usepackage[T1]{fontenc}
\usepackage{lmodern}
\usepackage[utf8]{inputenc} 
\usepackage{chemformula}
\usepackage{mathrsfs}
\usepackage{ntheorem}

\usepackage{longtable}
\usepackage{multirow}

\newcommand{\beq}{\begin{eqnarray} }
\newcommand{\eeq}{\end{eqnarray} }
\newcommand{\Beq}{\begin{eqnarray*} }
\newcommand{\Eeq}{\end{eqnarray*} }

\newcommand{\RNum}[1]{\uppercase\expandafter{\romannumeral #1\relax}}

\newcommand{\beginsupplement}{%
        \setcounter{table}{0}
        \renewcommand{\thetable}{S\arabic{table}}%
        \setcounter{figure}{0}
        \renewcommand{\thefigure}{S\arabic{figure}}%
        \setcounter{section}{0}
        \renewcommand{\thesection}{S\arabic{section}}%
        \setcounter{equation}{0}
        \renewcommand{\theequation}{S\arabic{equation}}%
     }

\begin{document}
\draft

\title{A Unified Symmetry Classification of Magnetic Orders via Spin Space Groups: Prediction of Coplanar Even-Wave Phases}
\author{Ziyin Song}
\affiliation{Beijing National Laboratory for Condensed Matter Physics,
    and Institute of Physics, Chinese Academy of Sciences, Beijing 100190, China}
\affiliation{University of Chinese Academy of Sciences, Beijing 100049, China}

\author{Ziyue Qi}
\affiliation{Beijing National Laboratory for Condensed Matter Physics,
    and Institute of Physics, Chinese Academy of Sciences, Beijing 100190, China}
\affiliation{University of Chinese Academy of Sciences, Beijing 100049, China}

\author{Chen Fang}
\affiliation{Beijing National Laboratory for Condensed Matter Physics,
    and Institute of Physics, Chinese Academy of Sciences, Beijing 100190, China}
    \affiliation{Kavli Institute for Theoretical Sciences, Chinese Academy of Sciences, Beijing 100190, China}

\author{Zhong Fang}
\affiliation{Beijing National Laboratory for Condensed Matter Physics,
    and Institute of Physics, Chinese Academy of Sciences, Beijing 100190, China}
\affiliation{University of Chinese Academy of Sciences, Beijing 100049, China}

\author{Hongming Weng}
\email{hmweng@iphy.ac.cn}
\affiliation{Beijing National Laboratory for Condensed Matter Physics,
    and Institute of Physics, Chinese Academy of Sciences, Beijing 100190, China}
\affiliation{Condensed Matter Physics Data Center, Chinese Academy of Sciences, Beijing 100190, China}

\begin{abstract}
    Spin space groups (SSGs) impose fundamentally different constraints on magnetic configurations in real and reciprocal spaces. As a consequence, the correspondence between real-space and momentum-space spin arrangements is far richer than traditionally assumed. Building on the complete enumeration of SSGs, we develop a systematic, symmetry-based framework that classifies all possible spin arrangements allowed by these groups. 
    This unified approach naturally incorporates conventional magnetic orders, altermagnetism, and $p$-wave magnetism as distinct symmetry classes. Crucially, our classification predicts a variety of novel magnetic phases, highlighted by the discovery of the coplanar even-wave magnet: a state that is non-collinear in real space but hosts a collinear even-wave spin polarization in $\mathbf{k}$-space. Analysis of a minimal model reveals that this phase is characterized by non-quantized spin polarization and exhibits a novel mechanism for symmetry-enforced zero polarization on non-degenerate bands. 
    Extending the framework from bulk crystals to layer SSGs appropriate for two-dimensional systems, we further predict layered counterparts and provide symmetry guidelines for designing bilayer coplanar odd-wave and even-wave magnets.
    We further validate this finding through first-principles calculations and propose \ch{CoCrO4} as a promising candidate for its experimental realization, thereby demonstrating the completeness and predictive power of the SSG-based classification of magnetic orders.
\end{abstract}

\maketitle

\section{Introduction}

Symmetry lies at the heart of condensed matter physics. For magnetic materials, the presence of ordered moments breaks time-reversal symmetry, and their crystal symmetry is commonly described by magnetic space groups (MSGs)\cite{Bradley1968,Lifshitz2004,Litvin2013MagneticGT}. 
However, in the weak spin-orbit coupling (SOC) regime relevant for many magnets, the spin and lattice degrees of freedom decouple to a good approximation, since the crystal-field splitting, electron-electron interactions, and kinetic hopping amplitudes are all typically much larger than SOC. In this limit, it is natural to regard spin rotations as independent of spatial operations, which gives rise to the more general framework of spin space groups (SSGs)\cite{Litvin1974,Brinkman1966,liu2022SSG}. Many magnetic and transport properties are therefore governed primarily by SSG symmetries, with SOC acting merely as a perturbation\cite{liu2025multipolar,liu2025SOM}.
Beyond these aspects, SSGs also refine the symmetry classification of band topology, enabling symmetry-enforced topological phases that are not captured by the conventional MSG description\cite{Yang2024SSGtopology,liu2022SSG,guo2021prlSSGtopology,InnovationSSGtopology,Corticelli2022SSGtopology,Chen2025magnons,Zhang2025IRSSG,Song2025SSGtopology}.

This SSG framework has recently attracted significant interest due to the discovery of altermagnetism\cite{smejkal2021altermagnetism,TJungwirth2022,mazin2022altermagnetism,Noda2016altermagnetism,smj2020altermagnetism,Ahn2019altermagnetism,Yuan2020altermagnetism,Bai2024altermagnetism}, $p$-wave magnetism\cite{hellenes2024pwavemagnets,brekke2024minimal,Yamada2025,Chakraborty2025,Sun2025pwave}, and other unconventional magnetic phases\cite{Liu2025unconventionalmagnetism,Chen2025magnons,yuan2025unconventionalmagnets,wu2004dynamic,wu2007fermiliquid,oddSSG}. Together with conventional ferromagnetism and antiferromagnetism, these phases exhibit distinct spin arrangements in real and reciprocal spaces, denoted by $S(\mathbf r)$ and $S(\mathbf k)$, respectively. Remarkably, altermagnets display zero net magnetization but exhibit spin splitting in reciprocal space, whereas $p$-wave magnetism, which is in fact one type of coplanar odd-wave magnetism\cite{oddSSG}, features coplanar spin arrangements in real space and collinear arrangements in reciprocal space. Despite their apparent differences, these magnetic phases share a common origin: their structures are fully dictated by how SSG operations constrain the spin configurations $S(\mathbf r)$ and $S(\mathbf k)$.

An SSG operation is written as $\{U_s||U_r\}$, where $U_s\in O(3)$ acts in spin space and $U_r$ is a real space operation containing both a rotational part and a translation.  
When $\det(U_s)=+1$, $U_s$ is a proper spin rotation. 
When $\det(U_s)=-1$, it is assumed that it contains time reversal symmetry $\mathcal{T}$ and is antiunitary, i.e., $U_s \sim \text{det}(U_s)U_s\cdot \mathcal{T}$.
The action of $\{U_s||U_r\}$ on the spin configuration in real and reciprocal spaces reads
\begin{equation}
\{U_s||U_r\} S(\mathbf r) = U_s S(U_r^{-1}\mathbf r),
\label{constrain real}
\end{equation}
\begin{equation}
\{U_s||U_r\} S(\mathbf k) = U_s S(\det(U_s)\,U_r^{-1}\mathbf k),
\label{constrain reciprocal}
\end{equation}
where the key distinction is the appearance of the factor $\det(U_s)$ in reciprocal space. This additional sign causes certain symmetry operations to act differently on $S(\mathbf r)$ and $S(\mathbf k)$, thereby allowing distinct combinations of real-space and momentum-space spin textures. Altermagnetism and $p$-wave magnetism arise precisely from this fundamental difference.

Recently, all SSGs have been enumerated\cite{xiao2023spin,jiang2023enumeration,ren2023enumeration}, providing a complete catalogue of possible symmetry groups. However, the implications of these groups for constraining spin configurations—particularly for relating $S(\mathbf r)$ and $S(\mathbf k)$ within a unified framework—remain underdeveloped.

In this work, we present a comprehensive analysis of how SSG symmetries constrain spin arrangements in real and reciprocal spaces. Within this symmetry-based framework, we show that the recently proposed altermagnetic and $p$-wave magnetic phases naturally emerge as distinct SSG symmetry classes. Moreover, we demonstrate that additional unconventional magnetic textures, including coplanar even-wave phase, are also predicted within this framework.

The paper is organized as follows. In Sec.~\ref{sec:ssg_constraints}, we analyze the SSG constraints on $S(\mathbf r)$ and $S(\mathbf k)$ for both three-dimensional bulk crystals and layered (two-dimensional) systems, and develop a general symmetry-based classification. Section~\ref{sec:model_d_wave} introduces a minimal model for the coplanar $d$-wave magnetic phase and compares it with existing magnetic classes. In Sec.~\ref{sec:material}, we propose a candidate material realization for this new phase. Finally, Sec.~\ref{sec:discussion} presents the discussion and concluding remarks.

\section{Symmetry Constraints on Spin Arrangements from Spin Space Groups}
\label{sec:ssg_constraints}

\begin{figure*}[!t]
    \centering
    \includegraphics[width=0.98\textwidth]{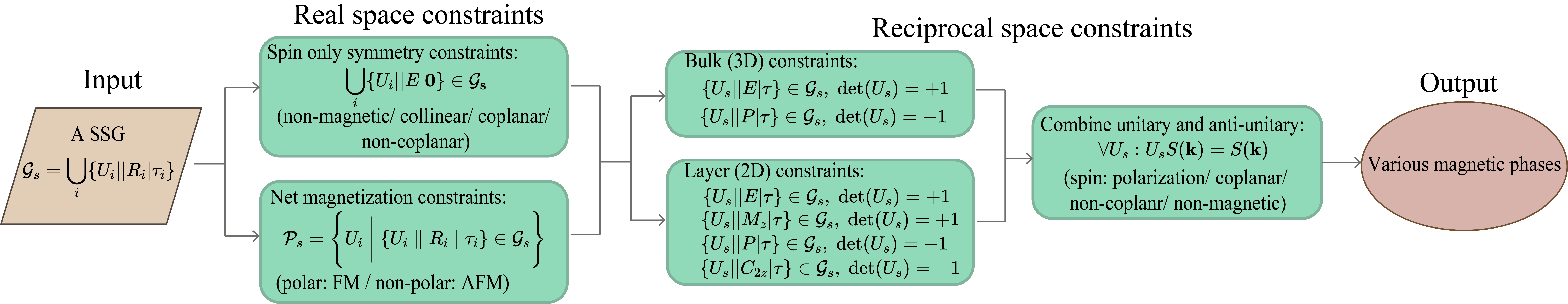}
    \caption{
    Flowchart illustrating the SSG-based classification of magnetic orders. 
    Starting from a given SSG $\mathcal{G}_s$, real space constraints (left) determine the local spin
    configuration and whether a net magnetization is allowed, while reciprocal space constraints (middle),
    formulated separately for bulk (3D) and layer (2D) systems, fix the allowed spin textures $S(\mathbf{k})$.
    Combining these operations yields the possible reciprocal space spin configurations
    (spin-polarized, coplanar, non-coplanar, or non-magnetic) and thus the corresponding magnetic phases.
    }
    \label{fig:flowchart}
\end{figure*}

In this section, we analyze how SSGs impose constraints on magnetic configurations in both real and reciprocal space. We show how these distinct constraints form the basis of a systematic classification of magnetic orders, as summarized in the flowchart in Fig.~\ref{fig:flowchart}.

We first examine the constraints imposed by SSGs on real-space magnetic configurations. 
These constraints fall into two distinct categories:

\begin{enumerate}

\item \textbf{Local onsite constraints:}  
If an onsite symmetry leaves an atomic position $\mathbf{r}$ invariant ($U_r^{-1}\mathbf{r}=\mathbf{r}$), it imposes a local condition on the magnetic moment,
\begin{equation}
    S(\mathbf{r}) = U_s\, S(\mathbf{r}),
    \label{eq:local_constraint}
\end{equation}
as expressed in Eq.~\ref{constrain real}.  
At a general Wyckoff position, only spin-only operations appear in the site symmetry. These can restrict the spins to be collinear or coplanar, whereas the presence of pure time-reversal symmetry forces $S(\mathbf{r})=0$, yielding a nonmagnetic configuration.

\item \textbf{Global constraints from the point group of the spin part:}  
The set of all spin rotations $\{U_s\}$ appearing in the SSG forms a point group in spin space (hereafter referred to as the point group of the spin part, $\mathcal{P}_s$, to avoid confusion with the spin point group).  
Whether this group admits a nonzero invariant vector determines whether a uniform magnetization is symmetry allowed: polar groups permit a net magnetic moment, whereas nonpolar groups forbid it.  
This provides the most general formulation of the FM/AFM dichotomy\cite{liu2025SOM}: polar $\mathcal{P}_s$ correspond to ferromagnetic-type orders, whereas nonpolar $\mathcal{P}_s$ correspond to antiferromagnetic-type orders.

\end{enumerate}

Combining these onsite and spin-part constraints yields a systematic classification of real-space spin arrangements, including nonmagnetic, collinear, coplanar, and non-coplanar configurations, with an additional distinction based on whether a net magnetization is allowed.

We now turn to reciprocal space, where the constraints imposed by SSGs become richer and often more decisive in determining the magnetic phase.  
At a general momentum $\mathbf{k}$, the spin texture $S(\mathbf{k})$ is constrained by the symmetries of its little group.  
These onsite symmetries may be unitary or antiunitary.  
The relevant unitary onsite operations take the general form $\{U_s || E \mid \tau\}$ with $\det(U_s)=+1$.  
Antiunitary onsite symmetries can appear as $\{U_s || P \mid \tau\}$ with $\det(U_s)=-1$, where the $\mathbf{k}$-inversion induced by time reversal is compensated by the spatial inversion $P$, making the operation onsite for $\mathbf{k}$.

Since these transformations impose local constraints on the spin texture, 
$S(\mathbf{k})$ must be a simultaneous invariant of all spin parts of the onsite symmetries:
\begin{equation}
\begin{aligned}
    S(\mathbf{k}) &= U_s\, S(\mathbf{k}), \\
    \{U_s || E \mid \tau\} \in \mathrm{SSG}, 
    &\qquad \det(U_s)=+1 \ (\text{unitary}), \\
    \{U_s || P \mid \tau\} \in \mathrm{SSG}, 
    &\qquad \det(U_s)=-1 \ (\text{antiunitary}).
\end{aligned}
\label{eq:kspace_constraint}
\end{equation}

By combining the reciprocal-space constraints with those in real space, 
we obtain a complete and systematic classification of magnetic orders.  
Building on the full enumeration of SSGs, this group-theoretical framework naturally reproduces all known collinear magnetic phases—including conventional antiferromagnetism and altermagnetism—while also revealing a broader landscape of coplanar orders.  
In particular, for coplanar SSGs, the reciprocal-space constraints admit a rich variety of spin textures, such as odd-wave magnetism\cite{oddSSG} and other nontrivial phases that we will discuss later.

\subsection{Collinear configurations}

A collinear SSG (1,421 in total) is defined by a spin-only subgroup that preserves a single axis (without loss of generality, we choose the $z$ axis). This subgroup contains a unitary rotation $C_{nz}$ and an antiunitary mirror $M_x$. The operation $C_{nz}$ protects the collinear configuration in both real and reciprocal spaces, while the antiunitary $M_x$ imposes an even-wave constraint on the spin texture:
\begin{equation}
    S^z(\mathbf{k}) = S^z(-\mathbf{k}),
\end{equation}
ensuring that the spin polarization is symmetric under $\mathbf{k} \rightarrow -\mathbf{k}$.

The classification of collinear SSGs is determined by their point group of the spin part, $\mathcal{P}_s$, together with the symmetry constraints at a generic $\mathbf{k}$ point in reciprocal space:
\begin{enumerate}
    \item $\mathcal{P}_s = 1$:  
    No SSG operation reverses the spin direction. The absence of spin-flipping symmetries permits a net magnetization, resulting in a collinear ferromagnet (FM) with an even-wave spin-polarized Fermi surface. This class contains 230 SSGs, identical to the number of space groups.

    \item $\mathcal{P}_s = m$ (generated by $M_z$):  
    The presence of $M_z$ (a spin-flip operation, equivalent to $C_{2x}$ or $C_{2y}$ in the presence of spin-only mirrors $M_x$ or $M_y$) forbids a net magnetization, leading to collinear antiferromagnetism (AFM). This AFM class further divides according to $\mathbf{k}$-space degeneracy:

    \begin{enumerate}
        \item \textbf{Conventional AFMs:}  
        These SSGs contain a degeneracy-enforcing constraint such as $\{M_z||P|\tau\}$ (an antiunitary operation involving inversion $P$) or $\{C_{2x(y)}||E|\tau\}$. These symmetries guarantee spin degeneracy at every $\mathbf{k}$ point, enforcing a vanishing reciprocal-space spin texture, $S(\mathbf{k}) = 0$. This class contains 769 SSGs.  
        We note that although unitary and antiunitary degeneracy-enforcing mechanisms differ in other physical consequences (e.g., Berry curvature\cite{LIU2022100343}), we do not distinguish them here since our focus is solely on the vanishing spin texture.

        \item \textbf{Altermagnets (AMs):}  
        These SSGs also have $\mathcal{P}_s = m$ (hence zero net magnetization) but lack the symmetries $\{M_z||P|\tau\}$ or $\{C_{2x(y)}||E|\tau\}$ that enforce degeneracy. Their absence allows a spin-split band structure in which the nonzero spin polarization $S(\mathbf{k})$ alternates across the Brillouin zone, producing the characteristic even-wave patterns ($d$-wave, $g$-wave, etc.). This class contains 422 SSGs.
    \end{enumerate}
\end{enumerate}

Thus, this SSG-based classification naturally divides collinear orders into FM and AFM categories, with the AFM category further subdivided into conventional AFMs and altermagnets according to their $\mathbf{k}$-space symmetry. This classification of collinear SSGs has been established in previous works\cite{xiao2023spin,jiang2023enumeration,ren2023enumeration}, and here we demonstrate how it follows directly from our general SSG-based framework.

\subsection{Coplanar configurations}

Coplanar SSGs (24,788 in total within $12\times$ supercell\cite{jiang2023enumeration}) are characterized by a spin-only antiunitary symmetry, $M_z$, which confines spins to the $xy$ plane. This $M_z$ symmetry has two key consequences. First, it dictates the wave parity of the $\mathbf{k}$-space spin texture:
\begin{equation}
     S(\mathbf{k}) =  M_z S(-\mathbf{k}).
     \label{eq:coplanar_wave}
\end{equation}
Since $M_z$ flips $S_z$ ($M_z S_z = -S_z$) but preserves $S_{x,y}$ ($M_z S_{x,y} = S_{x,y}$), this implies $S_z(\mathbf{k}) = -S_z(-\mathbf{k})$ (odd-wave) while $S_{x,y}(\mathbf{k}) = S_{x,y}(-\mathbf{k})$ (even-wave). Second, for every unitary operation $\{U_s||U_r|\tau\}$, a corresponding antiunitary counterpart $\{M_z U_s||U_r|\tau\}$ exists, enriching the $\mathbf{k}$-space constraints.

The point group of the spin part $\mathcal{P}_s$ for coplanar SSGs can be $C_{nz}$, $m$, or $C_{nv}$. The resulting $\mathbf{k}$-space spin textures are classified by their effective dimensionality (0D, 1D, 2D, or 3D) based on the onsite symmetry, such as $\{C_{nz}||E|\tau\}$, $\{C_{2x(y)}||E|\tau\}$, $\{M_z||P|\tau\}$, and $\{M_{x(y)}||P|\tau\}$.

\subsubsection{Non-magnetic (0D)}

A non-magnetic (0D) spin texture ($S(\mathbf{k}) = 0$) implies a fully spin-degenerate band structure. This is enforced by two primary mechanisms:
\begin{enumerate}
    \item \textbf{PT-like symmetry:} The presence of a $PT$-like symmetry, such as $\{C_{2z}||P|\tau\}$ (noting $C_{2z}M_z=-1$), enforces spin degeneracy at every $\mathbf{k}$ point (2,380 such SSGs).
    \item \textbf{Conflicting constraints:} A non-magnetic state can also arise from mutually contradictory constraints. For example, two orthogonal $C_2$ rotation onsite symmetry, $\{C_2||E|\tau_1\}$ and $\{C_2'||E|\tau_2\}$, may enforce $S(\mathbf{k})$ to lie along two different, perpendicular axes, for which the only solution is $S(\mathbf{k}) = 0$ (474 SSGs).
\end{enumerate}

\subsubsection{Spin-polarized (1D)}

A 1D (spin-polarized) $\mathbf{k}$-space texture occurs when SSG stabilizers constrain $S(\mathbf{k})$ to a single, fixed axis. This class reveals two physically distinct patterns:
\begin{enumerate}
    \item \textbf{$z$-axis polarization (odd-wave):} 
    The stabilizer operation $\{C_{nz}||E|\tau\}$ (with $n \ge 2$) constrains $S(\mathbf{k})$ to the $z$ axis. 
    Combined with Eq.~\eqref{eq:coplanar_wave}, this implies an odd-wave $z$-polarized texture
    \begin{equation}
        S^z(\mathbf{k}) = -S^z(-\mathbf{k}).
    \end{equation}
    Such coplanar odd-wave patterns were recently proposed in Ref.~\cite{oddSSG}, and in our enumeration, are realized by 17,830 distinct coplanar SSGs. 
    These SSGs can be further classified according to the symmetry of their spin polarization in reciprocal space into $p$-wave\cite{hellenes2024pwavemagnets} (17,049 SSGs), $f$-wave\cite{oddSSG} (637), $h$-wave (116), $j$-wave (14), and $l$-wave (14) subclasses. 
    The derivation of these wave patterns from the underlying SSG symmetries is given in the Supplementary Material~\ref{sm:wave pattern}.
    
    \item \textbf{In-plane polarization (Even-wave):} 
    The stabilizer $\{C_{2x(y)}||E|\tau\}$ constrains $S(\mathbf{k})$ to lie along the $x$ ($y$) axis, i.e., within the coplanar plane. From Eq.~\eqref{eq:coplanar_wave}, this in-plane polarization must be even-wave:
    \begin{equation}
        S^{x(y)}(\mathbf{k}) = S^{x(y)}(-\mathbf{k}).
    \end{equation}
    This pattern, featuring a symmetry-enforced even-wave spin polarization, is a distinct phase predicted by our classification. Our SSG enumeration shows that this class contains 517 polar SSGs, which can be understood as $s$-wave-like phases. The number 517 exactly matches that of type-IV MSGs, since such SSGs are generated by combining a pure space group with the spin translation operation $\{C_{2x(y)}||E|\tau\}$. In addition, there are 1,173 nonpolar SSGs in this class, among which 939 realize $d$-wave, 198 realize $g$-wave, and 36 realize $i$-wave even-spin textures. These numbers are exact within our classification scheme, where the SSG is restricted to be at a twofold supercell of the atomic space group.
    
    In Section~\ref{sec:model_d_wave} and the Supplementary Material, we will discuss in detail this coplanar even-wave phase (which includes the $\mathbf{k}$-space $d$-wave and higher-order moments), including its minimal model for different wave numbers and its many-to-one mapping with the altermagnetic phase.
\end{enumerate}

\subsubsection{Spin-coplanar (2D)}

A 2D (spin-coplanar) texture in $\mathbf{k}$ space is realized when $S(\mathbf{k})$ is constrained to a plane but not to a line. This arises from antiunitary mirror-like symmetries of the form $\{M||P|\tau\}$. We identify two types:
\begin{enumerate}
    \item \textbf{In-plane coplanar:} The operation $\{M_z||P|\tau\}$ (actually a pure inversion symmetry combined with the spin-only $M_z$) enforces $S(\mathbf{k})$ to lie within the $xy$-plane for all $\mathbf{k}$. This results in a coplanar texture in both real and reciprocal space (605 SSGs: 160 polar; 445 nonpolar).
    
    \item \textbf{Perpendicular-plane coplanar:} The operation $\{M_{x(y)}||P|\tau\}$ enforces $S(\mathbf{k})$ to lie within a plane perpendicular to the real-space spin plane (e.g., the $xz$- or $yz$-plane). This is a nontrivial case where the real-space and $\mathbf{k}$-space coplanar planes are mutually orthogonal (814 SSGs: 252 polar; 562 nonpolar).
\end{enumerate}

\subsubsection{Spin non-coplanar (3D)}

Finally, if an SSG lacks any of the $\mathbf{k}$-space onsite symmetries that constrain the dimensionality (0D, 1D, or 2D), the spin arrangement in reciprocal space is unconstrained. At a generic $\mathbf{k}$ point, $S(\mathbf{k})$ is free to point in any direction, resulting in a generic non-coplanar (3D) spin texture. This constitutes the most general case (995 SSGs: 262 polar; 733 nonpolar).

\subsection{Non-coplanar configurations}

Finally, we consider non-coplanar SSGs. These groups are defined by the absence of any non-trivial spin-only symmetry. Consequently, their real-space spin textures lack an intrinsic characteristic direction, such as the collinear axis or the coplanar plane found in the other classes.

Despite this lack of real-space constraint, the reciprocal-space spin texture $S(\mathbf{k})$ can still be highly constrained by $\mathbf{k}$-space stabilizers. Our SSG enumeration finds that these non-coplanar SSGs can produce a variety of $\mathbf{k}$-space spin arrangements, including:
\begin{enumerate}
    \item \textbf{Non-magnetic (0D):} A total of 12,838 SSGs enforce a non-magnetic $S(\mathbf{k})=0$ texture. As in the coplanar case, this is typically protected by a $PT$-like symmetry or by contradictory onsite symmetry constraints.
    
    \item \textbf{Spin-polarized (1D):} A total of 127,146 SSGs enforce a collinear $\mathbf{k}$-space spin texture, protected by an onsite symmetry operation such as $\{C_n || E | \tau\}$.
    
    \item \textbf{Spin-coplanar (2D):} A total of 8,323 SSGs enforce a coplanar $S(\mathbf{k})$ texture, protected by an onsite symmetry such as $\{M || P | \tau\}$.
    
    \item \textbf{Spin-non-coplanar (3D):} A total of 8,982 SSGs, which lack any of the $\mathbf{k}$-space onsite symmetry constraints required for 0D, 1D, or 2D arrangements.
\end{enumerate}
This demonstrates that our symmetry framework, based on the distinction between real-space and $\mathbf{k}$-space constraints, can be applied comprehensively to all magnetic classes, including this most general non-coplanar case.

\subsection{Extension to layered magnetic orders}
\label{sec:layered_orders}

In this subsection, we show that the unified symmetry classification of magnetic orders developed above can also be applied to layered magnetic orders, i.e., systems with an open boundary condition along one direction (taken to be the $z$ axis without loss of generality). For such layered structures, the appropriate symmetry group reduces from the 230 three-dimensional space groups to the 80 layer groups, and the corresponding SSG becomes a layer SSG constructed from these 80 layer groups\cite{layerSGs} and supercells along the $z$ axis are not allowed.

Our classification relies on real-space and reciprocal-space constraints imposed by the SSG. For layer SSGs, the real-space constraints remain unchanged, and magnetic orders can still be classified as non-magnetic, collinear, coplanar, or non-coplanar, with or without a net magnetization. However, in reciprocal space the constraints at a generic in-plane momentum $\mathbf{k}=(k_x,k_y)$ become richer. In particular, we now have
\begin{equation}
    \begin{aligned}
        S(\mathbf{k}) &= U_s\, S(\mathbf{k}), \\
        \{U_s || E \mid \tau\} \in \mathrm{SSG}, 
        &\qquad \det(U_s)=+1 \ \text{(unitary)}, \\
        \{U_s || M_z \mid \tau\} \in \mathrm{SSG}, 
        &\qquad \det(U_s)=+1 \ \text{(unitary)}, \\
        \{U_s || C_{2z} \mid \tau\} \in \mathrm{SSG}, 
        &\qquad \det(U_s)=-1 \ \text{(antiunitary)},\\
        \{U_s || P \mid \tau\} \in \mathrm{SSG}, 
        &\qquad \det(U_s)=-1 \ \text{(antiunitary)}.
    \end{aligned}
    \label{eq:kspace_constraint_layer}
\end{equation}
Within this framework, the dimensionality of the spin texture can be classified in exactly the same way as for three-dimensional SSGs. The conclusions of Sec.~\ref{sec:ssg_constraints} carry over essentially unchanged: one simply extends Eq.~\eqref{eq:kspace_constraint} by allowing the role of $\{U_s||E\mid\tau\}$ to be played by either $\{U_s||E\mid\tau\}$ or $\{U_s||M_z\mid\tau\}$, and similarly allowing $\{U_s||P\mid\tau\}$ to be replaced by either $\{U_s||P\mid\tau\}$ or $\{U_s||C_{2z}\mid\tau\}$. In particular, layer SSGs again support 0D, 1D, 2D, and 3D spin textures in $\mathbf{k}$ space. 

For collinear orders, this construction reproduces the family of layered altermagnets, which break spin-flip symmetry via the layer operations $\{M_z|\tau\}$, $\{P|\tau\}$, $\{E|\tau\}$, and $\{C_{2z}|\tau\}$. These four operations are precisely the layer counterparts generated by the replacements discussed above, and this family has been analyzed previously in Refs.~\cite{He2023layer,Qi2024layer,Tian2025layer}.

We highlight that this layer-based symmetry classification makes it particularly convenient to design coplanar odd-wave (including $p$-wave) magnets\cite{oddSSG} and coplanar even-wave magnets in bilayer structures. In the purely three-dimensional setting, a collinear spin texture in reciprocal space can only be protected by a symmetry of the form $\{C_2 || E \mid \tau\}$, i.e., a spin rotation combined with a pure translation, which can be rather rare in real materials. In contrast, in the layered setting one can construct a bilayer system from two coplanar layers that are related by a symmetry of the form $\{C_2 || M_z\}$, where the $C_2$ axis is either perpendicular or parallel to the coplanar plane. The former realizes a coplanar odd-wave magnetic phase, while the latter realizes a coplanar even-wave magnetic phase. 
This design principle is exploited explicitly in the Supplementary Material~\ref{sm:other even wave coplanar phase} and~\ref{i and g wave}, where we construct concrete bilayer models for even-wave phases.

\section{Model of Coplanar d-wave Magnetism}
\label{sec:model_d_wave}

\begin{table*}[!t]
    \centering
    \begin{tabular}{l|c|c|c}
     & \textbf{Collinear altermagnet} & \textbf{Coplanar odd-wave} & \textbf{Coplanar even-wave} \\ \hline
    \textbf{Spin polarization (k-space)} & 
    $\{C_{nz}||E\}$ & 
    {$\{C_{2z}||E|\tau\}$} &
    {$\{C_{2z}||E|\tau\}$} \\
    \hline
    \textbf{Collinear (r-space)} &
    $\{C_{nz}||E\}$ & 
    $\times$ & 
    $\times$ \\
    \hline
    \textbf{Coplanar (r-space)} & 
    $\times$ & 
    $\{M_{z}||E\}$ &
    $\{M_{x/y}||E\}$ \\
    \hline
    $\varepsilon_{\uparrow}(\mathbf{k}) = \varepsilon_{\downarrow}(-\mathbf{k})$ &
    $\times$ & 
    $\{M_{z}||E\}$ &
    $\times$ \\
    \hline
    $\varepsilon_{\uparrow}(\mathbf{k}) = \varepsilon_{\uparrow}(-\mathbf{k})$ &
    $\{M_{x,y}||E\}$ & 
    $\times$ &
    $\{M_{x/y}||E\}$ \\
    \hline
    \textbf{Spin polarization direction} &
    $S(\mathbf{k}) \parallel S(\mathbf{r})$ (z axis) &
    $S(\mathbf{k}) \perp S(\mathbf{r})$ &
    $S(\mathbf{k}) \parallel S(\mathbf{r})$ plane (yz/xz plane) \\
    \hline
    \textbf{odd-/even-wave} &
    $S(\mathbf{k}) = S(-\mathbf{k})$ &
    $S(\mathbf{k}) = -S(-\mathbf{k})$ &
    $S(\mathbf{k}) = S(-\mathbf{k})$  \\
    
    \end{tabular}
    \caption{Comparison of the symmetry properties of collinear altermagnets, coplanar odd-wave magnets, and coplanar even-wave magnetic orders. For clarity, we adopt a coordinate convention in which the spin polarization in momentum space is chosen along the $z$ axis for all three phases: the real-space moments are then collinear along $z$ for the altermagnet, coplanar in the $xy$ plane for the coplanar odd-wave phase, and coplanar in a plane containing the $z$ axis (e.g.\ $xz$ or $yz$) for the coplanar even-wave phase.}
    \label{table:spin_polarization_symmetry}
    \end{table*}

In this section, we present a minimal model for a coplanar $d$-wave magnetic phase. We highlight its critical differences from coplanar odd-wave magnetism, which can share a similar real-space spin configuration, and from $d$-wave altermagnetism, which exhibits a similar $\mathbf{k}$-space symmetry.
The construction of minimal models for other coplanar even-wave magnetic phases is presented in Supplementary Materials~\ref{sm:other even wave coplanar phase}.

\subsection{Symmetry and Model Hamiltonian}

\begin{figure*}[!t]
    \centering
    \includegraphics[width=0.98\textwidth]{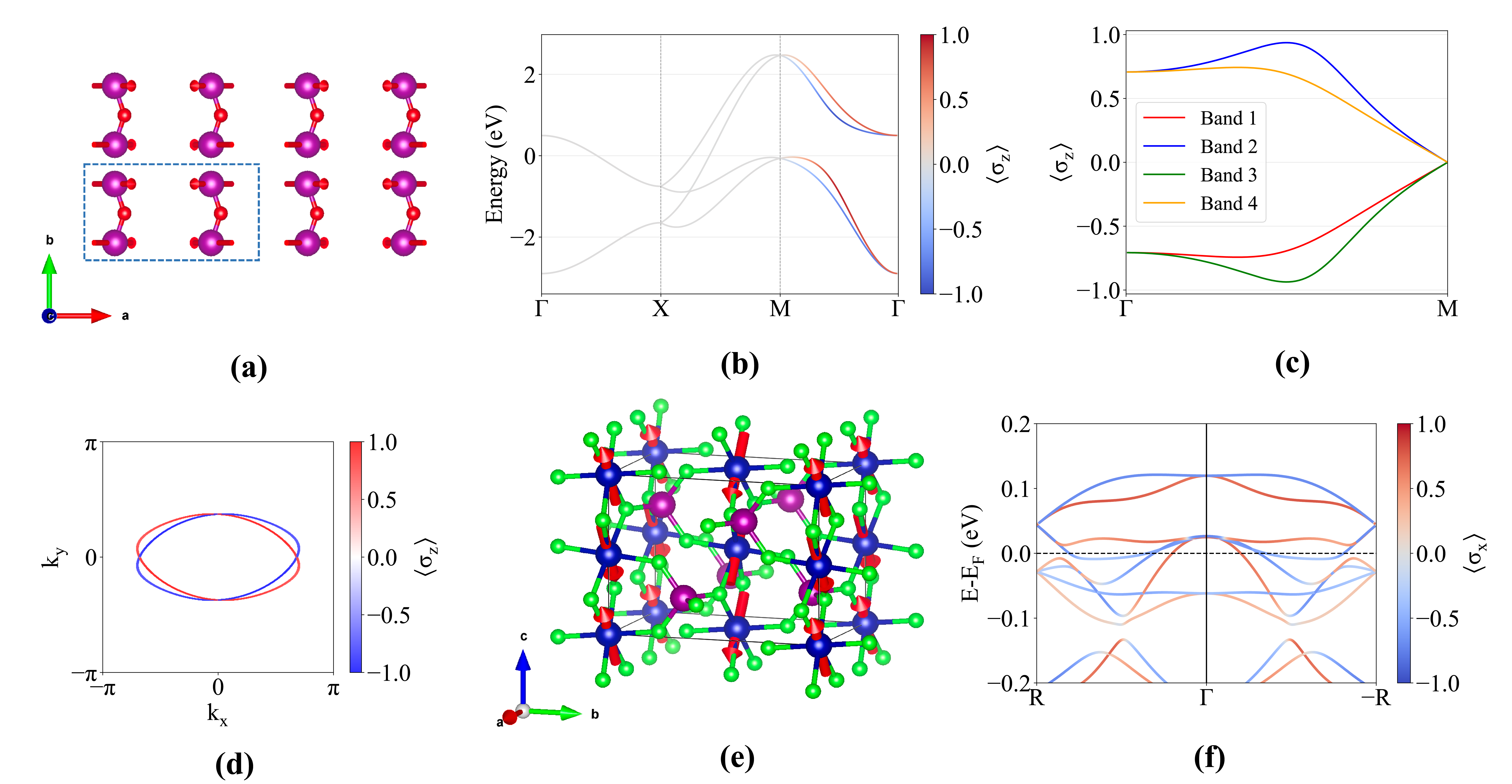}
    \caption{(\textbf{a}) Real-space lattice model for the coplanar $d$-wave magnet, fulfilling the required SSG symmetries. The dashed box indicates the magnetic unit cell. (\textbf{b}) Calculated band structure along high-symmetry lines and spin polarization ($\langle S_z \rangle$) projected onto the bands. (\textbf{c}) Spin polarization ($\langle S_z \rangle$) of four bands along the $\Gamma-M$ path. (\textbf{d}) Spin-polarized Fermi surface at $\mu = -2\,\text{eV}$, showing the characteristic $d$-wave anisotropy. (\textbf{e}) Magnetic structure of \ch{CoCrO4} exhibiting the coplanar magnetic configuration. (\textbf{f}) First-principles band structure of \ch{CoCrO4} and spin polarization ($\langle S_x \rangle$) projected onto the bands along the high-symmetry path.}
    \label{fig1:d-wave-coplanar}
\end{figure*}

We consider a model based on the real-space configuration shown in Fig.~\ref{fig1:d-wave-coplanar}(a), which features four magnetic atoms per unit cell. The magnetic moments are $(\alpha, 0, \beta)$, $(-\alpha, 0, \beta)$, $(\alpha, 0, -\beta)$, and $(-\alpha, 0, -\beta)$, respectively. This arrangement yields zero net magnetization and a spin configuration that is coplanar in the $xz$-plane, consistent with the spin-only antiunitary symmetry $\{M_{y}||E\}$. 

Despite the real-space coplanarity, the reciprocal-space spin texture is constrained by the SSG to be collinear and $z$-polarized. The symmetry constraints in the reciprocal space  are:
\begin{equation}
    \begin{aligned}
    &\{C_{2z} \| t\}:\quad S_x(k_x, k_y) = 0 ; \quad S_y(k_x, k_y) = 0\\
    &\{M_y \| E\}:\quad S_z(k_x,k_y) = S_z(-k_x, -k_y)\\  
    &\{C_{2x} \| M_y\}:\quad S_z(k_x,k_y) = -S_z(k_x, -k_y)
    \end{aligned}
    \end{equation}
The first constraint, $\{C_{2z} \| t\}$, enforces a pure $z$-axis spin polarization in $\mathbf{k}$-space. The second, $\{M_y \| E\}$, dictates an even-wave parity under inversion. The third, $\{C_{2x} \| M_y\}$, enforces an odd parity under reflection $k_y \to -k_y$. This momentum-space symmetry is precisely that of a $d$-wave state.

We then construct a minimal effective Hamiltonian for the nonmagnetic sites that respects these symmetries and the detailed derivation is given in Supplementary Material~\ref{sm:minimal_model}:
\begin{equation}
    \begin{aligned}
        \mathcal{H}(\mathbf{k}) & =2 t_{0}\cos \frac{k_{x}}{2} \sigma_{0} \tau_{1} + 2 t_{0}\cos k_{y} \sigma_{0} \tau_{0}  \\
        & +\left[t_{1} \cos k_{x}+t_{2} \cos k_{y}\right] \sigma_{1} \tau_{3} \\
        & +t_{3} \sin \frac{k_{x}}{2} \sigma_{1} \tau_{2} \\
        & +t_{4} \sin k_{y} \sigma_{2} \tau_{3}
        \end{aligned}
\end{equation}
where $\sigma_i$ and $\tau_i$ are Pauli matrices for the spin and sublattice degrees of freedom, respectively. The calculations are performed using the hopping parameters $t_0=-0.6$, $t_1=0.5$, $t_2=0.7$, $t_3=0.4$, and $t_4=0.6$.

The resulting band structure and spin polarization projected onto the bands are presented in Fig.~\ref{fig1:d-wave-coplanar}. Fig.~\ref{fig1:d-wave-coplanar}(b) shows the band structure along high-symmetry paths. Along the $\Gamma-X$ path, the bands are doubly degenerate and the total spin polarization $S_z(\mathbf{k})$ is zero, as the degeneracy is protected by the $\{C_{2x} || M_y\}$ symmetry. The $X-M$ path is non-degenerate, yet the expectation value of the spin polarization is still enforced to be zero by the $\{M_z || M_y\}$ symmetry. The $M-\Gamma$ path is non-degenerate and exhibits a finite spin polarization along the $z$-axis.

Fig.~\ref{fig1:d-wave-coplanar}(c) plots the expectation value of spin polarization $\langle S_z \rangle$ for these four bands along the $\Gamma-M$ path, where the expectation values of $\langle S_x \rangle$ and $\langle S_y \rangle$ are zero and therefore not shown. Finally, Fig.~\ref{fig1:d-wave-coplanar}(d) displays the spin-polarized Fermi surface at $\mu = -2\,\text{eV}$, which shows the characteristic $d$-wave anisotropy of this magnetic phase. 
Further analyses and model Hamiltonians for a coplanar $d$-wave phase with $C_4$ symmetry, as well as coplanar $g$- and $i$-wave phases, are presented in Supplementary Material~\ref{sm:other even wave coplanar phase} and~\ref{i and g wave}.

\subsection{Distinction from Other Phases}
\label{sec:distinction_from_other_phases}
In Table~\ref{table:spin_polarization_symmetry}, we compare the symmetry properties of collinear altermagnets, coplanar odd-wave, and coplanar even-wave magnetic orders.
For coplanar odd-wave magnetism, the $\mathbf{k}$-space spin polarization is perpendicular to the real-space spin plane, whereas for the coplanar even-wave, it is parallel to the real-space spin plane. Consequently, their respective spin-only symmetries result in different $\mathbf{k}$-space parities: coplanar odd-wave magnetism exhibits an odd-wave pattern, while the coplanar even-wave exhibits an even-wave pattern. Thus, the coplanar even-wave is a new phase that simultaneously shows similarities with collinear altermagnetism (even-wave parity) and coplanar odd-wave magnetism (coplanar in real space, collinear in $\mathbf{k}$-space).
We further show that coplanar even-wave SSGs admit a many-to-one mapping onto altermagnetic SSGs, as detailed in Supplementary Material~\ref{sm:many-to-one-mapping}.

It is crucial to emphasize another fundamental difference between this coplanar even-wave phase and collinear altermagnetism, even though both exhibit $z$-axis spin polarization and even-wave parity in $\mathbf{k}$-space. (This distinction also applies when comparing coplanar odd-wave magnetism with altermagnetism.)

In the collinear altermagnet, spin is a good quantum number. The Hamiltonian contains only $\sigma_0$ and $\sigma_3$ terms, and the eigenstates are pure spin-up or spin-down, i.e., $\langle \sigma_z \rangle = \pm 1$.

In our coplanar model (the same as in coplanar odd-wave magnetism), the Hamiltonian explicitly contains $\sigma_1$ and $\sigma_2$ terms. Consequently, spin is not a good quantum number, and the spin rotation symmetry $C_{nz}$ is broken. Although the spin expectation value $\vec{M} = (\langle \sigma_x \rangle, \langle \sigma_y \rangle, \langle \sigma_z \rangle)$ is polarized along $z$ (i.e., $M_x = M_y = 0$ and $M_z/|\vec{M}| = 1$ due to the $\{C_{2z} \| t\}$ symmetry), the local eigenstates are not pure spin states. As shown in Fig.~\ref{fig1:d-wave-coplanar}(c), the spin polarization varies continuously along the $\Gamma-M$ path, approaching zero smoothly as it reaches the $M$ point.

Moreover, this coplanar model exhibits a crucial distinction from collinear altermagnetism in how its spin polarization vanishes. In collinear altermagnets, $S_z(\mathbf{k})$ vanishes only on high-symmetry paths where spin degeneracy is protected. Our model shows this conventional behavior along the $\Gamma-X$ path [Fig.~\ref{fig1:d-wave-coplanar}(b)], where $S_z=0$ because the degeneracy is protected by the symmetry $\{C_{2x} || M_y\}$.

However, a novel mechanism is observed along the $X-M$ path. Here, the bands are clearly non-degenerate, yet the spin polarization $S_z(\mathbf{k})$ is still forced to be zero. This is a direct consequence of the onsite symmetry $\{M_z || M_y\}$. As this symmetry is anti-unitary and squares to $+1$ (not $-1$), it does not protect band degeneracy. Instead, it reverses the spin polarization operator ($S_z \to -S_z$), forcing its expectation value to vanish along the entire $X-M$ path. This explains why the spin polarization smoothly reaches zero precisely at the $M$ point, as observed in Fig.~\ref{fig1:d-wave-coplanar}(c).

\section{Material Realization: \ch{CoCrO4}}
\label{sec:material}

Our SSG-based framework provides a practical workflow for material classification: starting from an experimentally determined magnetic structure, one can identify its SSG and assign it to one of the symmetry classes in our scheme. 
By analyzing the magnetic structures in the MAGNDATA database\cite{Gallego2016magndataI}, we find 73 entries (including $\ch{CeNiAsO}$ proposed in Ref~\cite{hellenes2024pwavemagnets}) whose SSGs and magnetic configurations realize coplanar odd-wave magnets, defined as states with coplanar local moments, zero net magnetization, and a collinear spin polarization in momentum space perpendicular to the spin plane. 
In this search we did not require all magnetic atoms to be related by SSG operations (i.e., magnetically inequivalent sites are allowed). 
The MAGNDATA identifiers and chemical formulas of these 73 candidates are listed in the Supplementary Material~\ref{sm:p-wave magnet candidates}.

In addition, we identify cobalt chromate, \ch{CoCrO4}, as a promising candidate material for coplanar even-wave magnet. According to neutron diffraction studies, \ch{CoCrO4} is isostructural with \ch{CrVO4} and crystallizes in the orthorhombic space group \textit{Cmcm} (No. 63). The magnetic Co$^{2+}$ ions are located at the Wyckoff positions (0, 0, 0), (0, 0, 1/2), (1/2, 1/2, 1/2), and (1/2, 1/2, 0) within the unit cell\cite{Pernet1969,kume1971generation,Gallego2016magndataI}.

Crucially, the experimentally determined magnetic structure is non-collinear. As depicted in Fig.~\ref{fig1:d-wave-coplanar}(e), the arrangement of the magnetic $\text{Co}^{2+}$ spins (represented by the red arrows on the blue $\text{Co}$ atoms) is characterized by a combination of $A_x$ and $G_z$ modes. (The blue spheres represent Co atoms, the purple spheres represent Cr atoms, and the green spheres represent O atoms.) This superposition results in a coplanar spin texture where all magnetic moments lie in the $xz$-plane (at an angle of $55^\circ$ with the $x$-axis).
This real-space structure corresponds to a coplanar SSG whose spin-only point group is generated by $\{M_y||E\}$.

The full SSG is generated by operations including $\{E||P\}$, $\{C_{2y}||C_{2z}|0,0,\frac{1}{2}\}$, $\{E||C_{2x}\}$, and $\{C_{2x}|| E| \frac{1}{2} \frac{1}{2} 0\}$. 
This SSG falls into the coplanar $d$-wave class identified in Sec.~\ref{sec:ssg_constraints}. Specifically, the operation $\{C_{2x}|| E| \frac{1}{2} \frac{1}{2} 0\}$ constrains the reciprocal-space spin arrangement to be polarized along the $x$-axis ($S_y(\mathbf{k}) = S_z(\mathbf{k}) = 0$).

It is crucial to justify the use of the SSG framework here. First, this SSG is fully compatible with the non-magnetic space group (\textit{Cmcm}), as every spatial operation in \textit{Cmcm} is paired with a specific spin operation to form an element of the SSG. Second, the conventional Magnetic Space Group (MSG) for this structure is $Pbcn$ (BNS No. 60.417), which contains only one-fourth of the symmetry operations present in the full SSG. The MSG framework is therefore insufficient, and the SSG provides a richer, more accurate description. Finally, this approach is physically appropriate, as \ch{CoCrO4} is composed of 3d elements (Cr, Co) where SOC is weak, validating the decoupling of spin and spatial operations.

To confirm this prediction, we performed first-principles calculations and the computational details are given in Supplementary Material~\ref{sm:first-principles}. 
Fig.~\ref{fig1:d-wave-coplanar}(f) shows the calculated band structure and spin polarization of \ch{CoCrO4} along the $R$($0,\frac{\pi}{2},\frac{\pi}{2}$) to $\Gamma$($0,0,0$) to $-R$($0,-\frac{\pi}{2},-\frac{\pi}{2}$) path. The bands are clearly spin-polarized. As required by the SSG symmetry, the spin polarization is purely along the $x$-axis (polarization components along other axes are zero and not shown). Furthermore, the plot demonstrates the even-wave symmetry of this polarization, as $S_x(\mathbf{k}) = S_x(-\mathbf{k})$ (i.e., the polarization is identical on the $\Gamma \to R$ and $\Gamma \to -R$ segments). This even parity is a direct consequence of the spin-only symmetry $\{M_y||E\}$, as discussed in Sec.~\ref{sec:ssg_constraints}.

We also note that the spin polarization is separately forced to zero on the $k_y=0$ and $k_z=0$ planes. This is an additional constraint enforced by the liitle group symmetries $\{M_y||C_{2z}|0,0,\frac{1}{2}\}$ and $\{M_z||C_{2z}|0,0,\frac{1}{2}\}$, respectively.

\section{Conclusions and Outlook}
\label{sec:discussion}

In this work, we have developed a comprehensive and systematic classification of magnetic orders based on SSG symmetries. Our framework is built upon the fundamental principle that SSGs impose distinct constraints on spin configurations in real space ($S(\mathbf{r})$) and reciprocal space ($S(\mathbf{k})$). This discrepancy, which originates from the $\det(U_s)$ factor in the momentum-space transformation, allows for a rich variety of magnetic phases beyond conventional classifications.

Using the complete enumeration of SSGs, we have shown that our framework provides a unified understanding of known magnetic orders. It naturally incorporates conventional ferromagnetism and antiferromagnetism, as well as the recently discovered unconventional phases of altermagnetism and $p$-wave magnetism, as distinct symmetry classes.

More importantly, our systematic approach predicts a variety of novel magnetic states. We focused on a new predicted phase, the coplanar even-wave magnet, which is characterized by a coplanar spin arrangement in real space but hosts a collinear, even-wave spin polarization in reciprocal space. We presented a minimal model for this phase, highlighting its unique physical properties, such as the non-quantized nature of its spin polarization and a novel mechanism for symmetry-enforced zero polarization ($S_z(\mathbf{k})=0$) on non-degenerate bands. Furthermore, we identified the existing material \ch{CoCrO4}, with its coplanar magnetic structure, as a promising candidate for its experimental realization.

Looking forward, the classification framework presented here provides a powerful roadmap for the discovery and identification of new unconventional magnetic materials. While we focused on the coplanar even-wave phase, our analysis revealed many other exotic possibilities, particularly within the coplanar and non-coplanar SSGs (e.g., states with $\mathbf{k}$-space coplanar spin configuration), which represent a new frontier for theoretical and experimental exploration.

Investigating the physical consequences of these new magnetic orders is a critical next step. For instance, the unique interplay of non-collinear real-space spins and collinear, even-wave momentum-space spins in the coplanar even-wave phase may lead to novel magnetotransport and optical responses, distinct from both altermagnets and $p$-wave magnets. Finally, we call for further experimental investigation of our candidate material, \ch{CoCrO4}, using techniques such as spin-polarized ARPES, to directly visualize the predicted $d$-wave spin polarization in its Fermi surface and confirm its status as the first coplanar even-wave magnet.

\acknowledgments{
    This work was supported by the Science Center of the National Natural Science Foundation of China (Grants No. 12188101 and No. 12325404), the National Key R\&D Program of China (Grants No. 2022YFA1403800, No. 2023YFA1607400, No. 2024YFA1408400 and No. 2023YFA1406704), the National Natural Science Foundation of China (Grants No. 12274436, No. 11925408, No. 11921004), and H.W. acknowledges support from the New Cornerstone Science Foundation through the XPLORER PRIZE.
    }

\normalem
\bibliography{ref}

\clearpage
\newpage
\onecolumngrid

\begin{center}
\textbf{\center{\large{Supplementary Material}}}
\end{center}

\beginsupplement

\tableofcontents

\section{Derivation of the Minimal Model}\label{sm:minimal_model}

In this section, we show how to derive the minimal tight-binding Hamiltonian for the coplanar $d$-wave magnetic phase. The model is constructed on a two-sublattice (A and B) basis, spanned by the direct product of spin and sublattice degrees of freedom: $\Psi = (c_{A\uparrow}, c_{A\downarrow}, c_{B\uparrow}, c_{B\downarrow})^T$.

The minimal model must satisfy the following three SSG symmetry constraints that define the coplanar $d$-wave phase (as analyzed in the main text):
\begin{equation}
    \begin{aligned}
    &\{C_{2z} \| t\}:\quad S_x(\mathbf{k}) = 0; \quad S_y(\mathbf{k}) = 0 \quad (\text{Collinear } S_z \text{ polarization in } \mathbf{k}\text{-space})\\
    &\{M_y \| E\}:\quad S_z(\mathbf{k}) = S_z(-\mathbf{k}) \quad (\text{Even-wave parity})\\  
    &\{C_{2x} \| M_y\}:\quad S_z(\mathbf{k}) = -S_z(\mathbf{k}_x, -\mathbf{k}_y) \quad (\text{Symmetry for } d\text{-wave shape})
    \end{aligned}
    \label{eq:ssg_constraints}
\end{equation}

The SSG operations that generate these constraints are represented by the following matrices in the $(\sigma_i \otimes \tau_j)$ basis (where $\sigma$ and $\tau$ denote spin and sublattice degrees of freedom, respectively):

\begin{table}[h!]
    \centering
    \setlength{\tabcolsep}{10pt}
    \renewcommand{\arraystretch}{1.3}
    \begin{tabular}{c|c|c|c}
    \hline
    \textbf{Generator} & $\{C_{2z} \| t\}$ & $\{M_y \| E\}$ & $\{C_{2x} \| M_y\}$ \\
    \hline
    \textbf{Matrix $\mathcal{U}$} & $-i\sigma_3 \tau_1$ & $\sigma_0 \tau_0 \kappa$ & $-i\sigma_1 \tau_0$ \\
    \hline
    \textbf{$\mathbf{k}$ change} & $\mathbf{k} \to \mathbf{k}$ & $\mathbf{k} \to -\mathbf{k}$ & $\mathbf{k} \to (k_x, -k_y)$ \\
    \hline
    \end{tabular}
    \caption{Operators of the SSG generators in the basis of the Hamiltonian, and the corresponding change in momentum $\mathbf{k}=(k_x, k_y)$.}
    \label{smtab:generators}
\end{table}

The Hamiltonian $H(\mathbf{k})$ must be invariant under all SSG operations $\mathcal{U}$:
\begin{equation}
    H(\mathbf{k}) = \mathcal{U} H(\mathbf{k}') \mathcal{U}^{-1},
    \label{eq:invariance}
\end{equation}
where $\mathbf{k}'$ is the momentum after the operation (e.g., $\mathbf{k}' = -\mathbf{k}$ for $\{M_y \| E\}$). For anti-unitary operations ($\mathcal{U} = A \kappa$), the invariance requires $H(\mathbf{k}) = A H(\mathbf{k}')^* A^{-1}$.

We systematically consider terms of the form $\sigma_i \tau_j f(\mathbf{k})$ that survive these constraints. Due to the minimal model assumption, we only consider $\sin k_i$ and $\cos k_i$ dependencies ($\frac{k_x}{2}$ with $\tau_1$ and $\tau_2$).

\textbf{Constraints Imposed by $\{C_{2z} \| t\}$:}

The operation is unitary ($\mathcal{U} = -i\sigma_3 \tau_1$) and leaves $\mathbf{k}$ invariant. The invariance condition $H(\mathbf{k}) = \mathcal{U} H(\mathbf{k}) \mathcal{U}^{-1}$ requires that $H(\mathbf{k})$ commutes with $\mathcal{U}$: $[\mathcal{U}, H(\mathbf{k})] = 0$.

\begin{itemize}
    \item Surviving Terms (Commuting): $\sigma_0 \tau_0, \sigma_0 \tau_1, \sigma_1 \tau_2, \sigma_1 \tau_3, \sigma_2 \tau_2, \sigma_2 \tau_3, \sigma_3 \tau_0, \sigma_3 \tau_1$.
    \item Forbidden Terms (Anti-Commuting): $\sigma_0 \tau_2, \sigma_0 \tau_3, \sigma_1 \tau_0, \sigma_1 \tau_1, \sigma_2 \tau_0, \sigma_2 \tau_1, \sigma_3 \tau_2, \sigma_3 \tau_3$. 
\end{itemize}

\textbf{Constraints Imposed by $\{M_y \| E\}$:}

This operation is anti-unitary ($\mathcal{U} = \sigma_0 \tau_0 \kappa$) and transforms $\mathbf{k} \to -\mathbf{k}$. The invariance condition $H(\mathbf{k}) = A H(-\mathbf{k})^* A^{-1}$ simplifies to $H(\mathbf{k}) = H(-\mathbf{k})^*$.

For simplicity, we test one of the surviving terms, $H_1(\mathbf{k}) = f(\mathbf{k}) \sigma_2 \tau_3$.
\begin{enumerate}
    \item The condition is $f(\mathbf{k}) \sigma_2 \tau_3 = (f(-\mathbf{k}) \sigma_2 \tau_3)^* = f(-\mathbf{k}) (\sigma_2 \tau_3)^*$.
    \item Since $\sigma_2 \tau_3$ is a purely imaginary matrix, $(\sigma_2 \tau_3)^* = -\sigma_2 \tau_3$.
    \item The condition becomes $f(\mathbf{k}) \sigma_2 \tau_3 = f(-\mathbf{k}) (-\sigma_2 \tau_3)$, which implies $f(\mathbf{k}) = -f(-\mathbf{k})$.
\end{enumerate}
This requires $f(\mathbf{k})$ to be an \textbf{odd} function of $\mathbf{k}$ (e.g., $\sin k_x$ or $\sin k_y$).

\textbf{Constraints Imposed by $\{C_{2x} \| M_y\}$:}

This final operation is unitary ($\mathcal{U} = -i\sigma_1 \tau_0$) and transforms $\mathbf{k} \to (k_x, -k_y)$. We now test the terms that survived the first two constraints, such as $H_1(\mathbf{k}) = f(\mathbf{k}) \sigma_2 \tau_3$ where $f(\mathbf{k})$ must be odd.
\begin{enumerate}
    \item The invariance condition is $H_1(\mathbf{k}) = \mathcal{U} H_1(k_x, -k_y) \mathcal{U}^{-1}$.
    \item The matrix $\sigma_2 \tau_3$ anti-commutes with $\mathcal{U} = -i\sigma_1 \tau_0$, meaning $\mathcal{U} \sigma_2 \tau_3 \mathcal{U}^{-1} = -\sigma_2 \tau_3$.
    \item The condition becomes $f(k_x, k_y) \sigma_2 \tau_3 = f(k_x, -k_y) (-\sigma_2 \tau_3)$, which implies $f(k_x, k_y) = -f(k_x, -k_y)$.
\end{enumerate}
Therefore, the only surviving term of this form is $\sin k_y \sigma_2\tau_3$.

Following this systematic constraint analysis for all surviving matrix bases, we can construct the full minimal model for the coplanar $d$-wave magnetic phase as presented in the main text.

\section{Details of First-Principles Calculations}\label{sm:first-principles}

The first-principles calculations were performed using the Vienna ab initio simulation package (VASP), utilizing the generalized gradient approximation (GGA) of the Perdew-Burke-Ernzerhof (PBE) type for the exchange-correlation potential. We explicitly did not take spin-orbit coupling (SOC) into account, as its minor influence on the band structure does not alter the symmetry-derived conclusion, and its inclusion often complicates the convergence process.

In the self-consistent field (SCF) calculation, a Monkhorst-Pack ($5 \times 5 \times 5$) k-point mesh and an energy cutoff of 600 eV were used. The self-consistent iterations converged reliably, achieving an energy difference criterion within $1\times 10^{-6}\,\text{eV}$. Constraints on the magnetic moments were not applied, allowing the magnetic configuration to be determined purely by the self-consistent potential.

The resulting local magnetic moment of the Co ion converged to a magnitude of $2.384\,\mu_B$, with the moment direction lying at an angle of $58.55^\circ$ with respect to the $x$-axis. This calculated moment is in very good agreement with experimental results, demonstrating the robustness and reasonableness of the material structure used for our magnetic model.

\section{Many-to-One Mapping from Coplanar Even-Wave to Altermagnetic Phase}\label{sm:many-to-one-mapping}

In this section, we demonstrate a formal many-to-one mapping between the SSGs of the coplanar even-wave phase and those of the altermagnetic (collinear) phase. This mapping is based on the key insight that both phases can share an identical set of reciprocal-space symmetry constraints, even though their real-space spin arrangements are fundamentally different (coplanar vs. collinear).

In our definition, the coplanar even-wave phase is characterized by: (1) An anti-unitary spin-only symmetry defining the plane (e.g., $\{M_y||E\}$ for the $xz$-plane). (2) An onsite symmetry (e.g., $\{C_2||\tau\}$) that constrains the $\mathbf{k}$-space spin polarization to an axis parallel to that plane.

A corresponding collinear SSG can be constructed by formally mapping the generators. The unitary onsite symmetry that protects the in-plane polarization in the coplanar case (e.g., $\{C_2||\tau\}$) plays the same functional role as the unitary spin-only rotation ($\{C_{nz}||E\}$) that defines the axis in the collinear case. The anti-unitary spin-only operation (e.g., $\{M_y||E\}$) dictates the even-wave parity in both phases.

This mapping can be illustrated using the generators of our coplanar $d$-wave model (LHS) and their counterparts in a corresponding collinear SSG (RHS):
\begin{enumerate}
    \item $\{C_{2z} \| t\}$ (Coplanar stabilizer protecting $S_z$) $\to$ $\{C_{nz} \| E\}$ (Collinear spin-only op protecting $S_z$)
    \item $\{M_y \| E\}$ (Coplanar spin-only, even-wave parity) $\to$ $\{M_y||E\}$ (Collinear spin-only, even-wave parity)
    \item $\{C_{2x} \| M_y\}$ (Coplanar $\mathbf{k}$-space constraint) $\to$ $\{C_{2x} \| M_y\}$ (Collinear $\mathbf{k}$-space constraint)
\end{enumerate}
One can verify that the set of generators on the right-hand side, which has the same $\mathbf{k}$-space symmetry actions as the left, constitutes a valid collinear SSG.

This mapping can also be understood using the International Notation. An example of a coplanar even-wave SSG is ${C}^{1}{m}^{2_{010}}{c}^{2_{010}}{m}|(1,1,1;2_{100})^m1$. Here, $2_{100}$ constrains the $x$-polarization, and $2_{010}$ acts as a spin-flip. This SSG maps to the collinear SSG ${C}^{1}{m}^{-1}{c}^{-1}{m}^{\infty m}1$, which notably lacks $PT$ symmetry and is therefore an altermagnetic phase. The translational parts of the operators can be omitted in this mapping because the definition of this coplanar phase already requires that they do not flip the spin, simplifying the correspondence.

Using this mapping scheme, one can unambiguously find a corresponding collinear (altermagnetic) SSG for every coplanar even-wave SSG. However, the mapping from the altermagnetic phase back to the coplanar even-wave phase is not a simple one-to-one correspondence. Because this mapping is defined purely by $\mathbf{k}$-space symmetry, a single altermagnetic SSG may correspond to multiple, distinct coplanar SSGs, or none at all.

This one-to-many correspondence arises because different operations can be functionally equivalent in the collinear case but non-equivalent in the coplanar case. For example, two distinct spin operations involving $C_{2x}$ and $C_{2y}$ might both act as simple spin-flips on a $z$-polarized collinear state. However, these operations are non-equivalent within the coplanar spin group, thus generating two distinct coplanar SSGs that map to the same altermagnetic SSG.

The one-to-none correspondence occurs when an altermagnetic SSG is built from a non-symmorphic space group that is incompatible with the definition of a coplanar even-wave phase. For instance, the space group $P2_1 2_1 2_1$ (No. 19) can form an altermagnetic collinear SSG ($P^{-1} 2_1 \;^{1}2_1 \;^{1}2_1$). However, the structure of its non-symmorphic operations cannot be paired with the spin operations required to satisfy the definition of a coplanar even-wave SSG, resulting in an altermagnetic phase with no coplanar counterpart.

\section{d-wave model with C4 symmetry}\label{sm:other even wave coplanar phase}

\begin{figure*}[!t]
    \centering
    \includegraphics[width=0.98\textwidth]{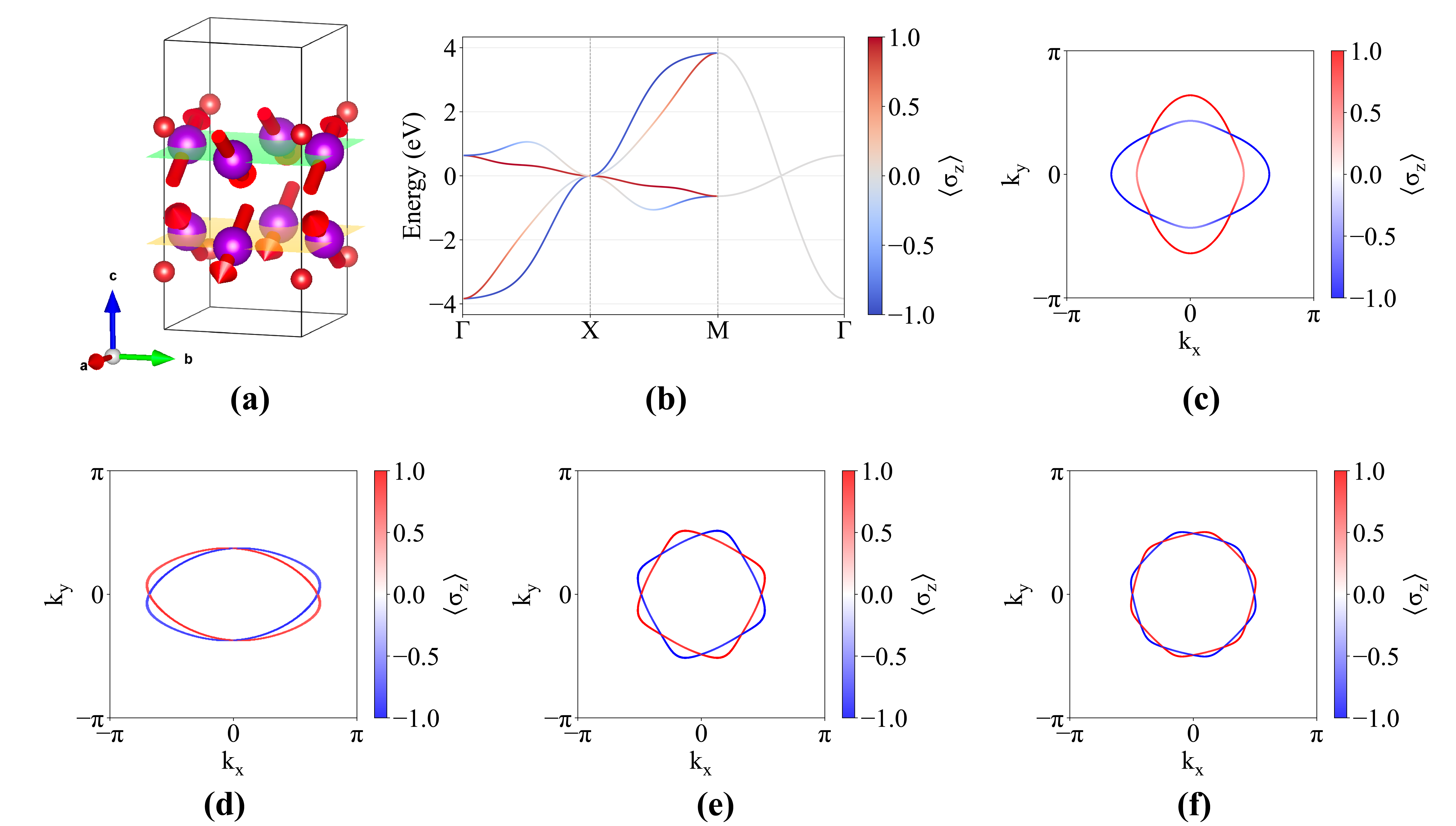}
    \caption{(\textbf{a}) Real-space lattice model for the coplanar $d$-wave magnet with $C_4$ symmetry, fulfilling the required SSG symmetries. The green and yellow planes show two layers of magnetic atoms related by the SSG symmetry $\{C_{2z}||M_z\}$. (\textbf{b}) Calculated band structure along high-symmetry lines and spin polarization $\langle S_z \rangle$ projected onto the bands for the $C_4$ model. (\textbf{c}) Spin-polarized Fermi surface for the $C_4$ model at $\mu = -2\,\text{eV}$, showing the characteristic $d$-wave anisotropy compatible with $C_4$ symmetry. (\textbf{d, e, f}) Schematics of various even-wave spin-polarized Fermi surface patterns: (\textbf{d}) a $d$-wave pattern (corresponding to the main text model), (\textbf{e}) a $g$-wave pattern, and (\textbf{f}) an $i$-wave pattern, generated by models detailed in the Supplementary Material.}
    \label{smfig:other even wave coplanar phase}
\end{figure*}

In the main text, we have shown a minimal model for the coplanar $d$-wave phase, and here we show another minimal model for the coplanar $d$-wave phase with $C_4$ symmetry. This model is based on a layered structure, as illustrated in Fig.~\ref{smfig:other even wave coplanar phase}(a), where two layers are related by the SSG operations $\{C_{2z}||M_z\}$.

We consider a two-layer, two-sublattice model (four-band model), with the following SSG symmetry constraints:
\begin{equation}
    \begin{aligned}
    &\{C_{2z} \| M_z\}:\quad S_x(k_x, k_y) = 0 ; \quad S_y(k_x, k_y) = 0\\
    &\{M_y \| E\}:\quad S_z(k_x,k_y) = S_z(-k_x, -k_y)\\  
    &\{C_{2x} \| C_{4z}\}:\quad S_z(k_x,k_y) = -S_z(-k_y, k_x)\\
    &\{E \| M_x\}:\quad S_z(k_x,k_y) = S_z(-k_x, k_y)
    \end{aligned}
    \end{equation}

Similar to the analysis in Sec.~\ref{sm:minimal_model}, the SSG generators act on the Hamiltonian basis as summarized in Table~\ref{smtab:generatorsC4}:
\begin{table}[h!]
    \centering
    \setlength{\tabcolsep}{10pt}
    \renewcommand{\arraystretch}{1.3}
    \begin{tabular}{c|c|c|c|c}
    \hline
    \textbf{Generator} & $\{C_{2z} \| M_z\}$ & $\{M_y \| E\}$ & $\{C_{2x} \| C_{4z}\}$ & $\{E \| M_x\}$ \\
    \hline
    \textbf{Matrix $\mathcal{U}$} & $-i\sigma_3 \tau_1$ & $\sigma_0 \tau_0 \kappa$ & $-i\sigma_1 \tau_0$ &  $\sigma_0\tau_0$\\
    \hline
    \textbf{$\mathbf{k}$ change} & $\mathbf{k} \to \mathbf{k}$ & $\mathbf{k} \to -\mathbf{k}$ & $\mathbf{k} \to (-k_y, k_x)$ &$\mathbf{k} \to (-k_x, k_y)$ \\
    \hline
    \end{tabular}
    \caption{Operators of the SSG generators in the basis of the Hamiltonian, and the corresponding change in momentum $\mathbf{k}=(k_x, k_y)$.}
    \label{smtab:generatorsC4}
\end{table}

The resulting minimal Hamiltonian $\mathcal{H}(\mathbf{k})$ that satisfies all the above SSG constraints is:

\begin{equation}
    \begin{aligned}
        \mathcal{H}(\mathbf{k}) & = t_{0}(\cos k_{x} + \cos k_y)\sigma_0 \tau_0 \\
        & +t_{1} (\cos k_{x} + \cos k_y) \sigma_{0} \tau_{1} \\
        & +t_{2} (\cos k_{x} + \cos k_y) \sigma_{1} \tau_{3} \\
        & +t_{3} \sin (k_x+k_y) \sin (k_x-k_{y}) \sigma_{2} \tau_{2}\\
        & +t_{4} \sin (k_x+k_y) \sin (k_x-k_{y}) \sigma_{3} \tau_{0}\\
        & +t_{5} \sin (k_x+k_y) \sin (k_x-k_{y}) \sigma_{3} \tau_{1}\\
    \end{aligned}
\end{equation}

Here we must introduce the second order of $\mathbf{k}$. Since it is not sufficient to describe the symmetry in this four bands model if only with the first order of $\mathbf{k}$.
Figure~\ref{smfig:other even wave coplanar phase}(b) and (c) show the resulting spin-polarized band structure and Fermi surface for parameters $t_0, t_1, t_2, t_3, t_4, t_5 = -0.8, -1, 0.5,0.8,0.6,0.5$. The results confirm the expected even-wave spin polarization and the characteristic d-wave anisotropy under $C_4$ symmetry.

\section{Models for the coplanar g- and i-wave phases}
\label{i and g wave}

In altermagnetic systems, the magnetic order may exhibit higher even-parity form factors such as $g$- and $i$-wave patterns. In this section, we construct explicit tight-binding models realizing coplanar spin structures with $g$- and $i$-wave symmetry, and we present their spin-polarized Fermi surfaces, which display the characteristic symmetry patterns dictated by the corresponding SSGs.

\textbf{Coplanar $g$-wave phase}

For the $g$-wave coplanar state, we consider the SSG generated by  $\{C_{2z}||M_z\}$, $\{M_y||E\}$, $\{E||C_{4z}\}$ and $\{C_{2z}||M_x\}$. 
A minimal four-band Hamiltonian respecting these generators is
\begin{equation}
    \begin{aligned}
        \mathcal{H}(\mathbf{k}) & = t_{0}(\cos k_{x} + \cos k_y)\sigma_0 \tau_0 \\
        & +t_{1} (\cos k_{x} + \cos k_y) \sigma_{0} \tau_{1} \\
        & +t_{2} (\cos k_{x} + \cos k_y) \sigma_{1} \tau_{3} \\
        & +t_{3} f_1(k_x, k_y) \sigma_{2} \tau_{2}\\
        & +t_{4} f_1(k_x, k_y) \sigma_{3} \tau_{0}\\
        & +t_{5} f_1(k_x, k_y) \sigma_{3} \tau_{1}\\
    \end{aligned}
\end{equation}
where $f_1(k_x,k_y) = \sin k_x \sin k_y (\cos k_x - \cos k_y)$ is chosen to satisfy the symmetry constraints $f_1(k_x, k_y) = -f_1(-k_x, k_y)$ and $f_1(k_x, k_y) = -f_1(-k_y, k_x)$. 

\textbf{Coplanar $i$-wave phase}

For the coplanar $i$-wave phase, we consider the SSG generated by
$\{C_{2z}||M_z\}$, $\{M_y||E\}$, $\{E||C_{6z}\}$ and $\{C_{2z}||M_x\}$. 
Here the model is naturally expressed in the hexagonal reciprocal basis $(k_1,k_2)$. The four-band Hamiltonian takes the form
\begin{equation}
    \begin{aligned}
        \mathcal{H}(\mathbf{k}) & = t_{0}f_2(k_1, k_2)\sigma_0 \tau_0 \\
        & +t_{1} f_2(k_1, k_2) \sigma_{0} \tau_{1} \\
        & +t_{2} f_2(k_1, k_2) \sigma_{1} \tau_{3} \\
        & +t_{3} f_3(k_1, k_2) \sigma_{2} \tau_{2}\\
        & +t_{4} f_3(k_1, k_2) \sigma_{3} \tau_{0}\\
        & +t_{5} f_3(k_1, k_2) \sigma_{3} \tau_{1}\\
    \end{aligned}
\end{equation}
where the factors are
\begin{equation}
    \begin{aligned}
        f_2(k_1, k_2) &= \cos k_1+\cos k_2 + \cos(k_1+k_2)\\
        f_3(k_1, k_2) &= \left[\cos k_1 - \cos(k_1 + k_2)\right]\times(\cos k_2 -\cos k_1) \times (\cos(k_1+k_2) - \cos k_2)
    \end{aligned}
\end{equation}

Figures~\ref{smfig:other even wave coplanar phase}(e) and (f) display the resulting spin-polarized Fermi surfaces for the parameter set $t_0, t_1, t_2, t_3, t_4, t_5 = -0.8, -1, 0.5,0.8,0.6,0.5$. 
The emergent Fermi-surface textures exhibit the characteristic $g$- and $i$-wave symmetry patterns analogous to those of the corresponding altermagnetic states. Together with Figs.~\ref{smfig:other even wave coplanar phase}(c) and (d), these models provide explicit realizations of coplanar magnetic textures protected by distinct SSG symmetries.

\section{Determination of the wave pattern for spin-polarized phases}
\label{sm:wave pattern}

In this section, we explain how to determine the precise wave character of a spin-polarized phase, i.e., whether its $\mathbf{k}$-space spin texture belongs to the $p$-, $f$-, $h$-, $\dots$ family for odd-wave phases, or to the $d$-, $g$-, $i$-wave family for even-wave phases.

The $s$-wave case is special: an $s$-wave spin texture is constant in $\mathbf{k}$ space and therefore does not require vanishing net magnetization. It corresponds to a ferromagnetic-type configuration, such as a collinear ferromagnet or the polar coplanar even-wave phase discussed in the main text.

For all higher waves, the wave pattern characterizes the symmetry of the spin polarization in reciprocal space. We focus on symmetry-enforced spin-polarized phases, where the spin at a generic $\mathbf{k}$ point is constrained to lie along a fixed axis (e.g., the $z$ axis). In this case the spin polarization can be treated as a scalar function $s(\mathbf{k})$. The symmetry operations $\{U_s||R|\tau\}$ of the SSG act on this function as
\begin{equation}
    s(\mathbf{k}) =
    \begin{cases}
      +\,s(R^T\mathbf{k}), & \text{for $U_s$ preserve the spin direction}, \\
      -\,s(R^T\mathbf{k}), & \text{for $U_s$ that flip the spin direction},
    \end{cases}
\end{equation}
so $s(\mathbf{k})$ carries a one-dimensional representation of the SSG in which spin-preserving operations have character $+1$ and spin-flipping operations have character $-1$.

As a concrete example, consider the $d$-wave coplanar model in the main text. The spin-only antiunitary generator $\{M_y||E\}$ preserves the spin direction while inverting momentum, so it enforces
\(
s(\mathbf{k}) = +\,s(-\mathbf{k}).
\)
On the other hand, the generator $\{C_{2x}||M_y\}$ flips the spin and maps $\mathbf{k}\to(k_x,-k_y)$, which enforces
\(
s(\mathbf{k}) = -\,s(k_x,-k_y).
\)
The monomial $k_x k_y$ satisfies both constraints,
\(
k_x k_y = (+1)\,(-k_x)(-k_y) = (-1)\,k_x(-k_y),
\)
and therefore transforms according to exactly this one-dimensional representation. This is what we mean when we say that the spin texture has $d$-wave symmetry.

More generally, we define the wave number $l$ as the degree of the polynomial in $(k_x,k_y,k_z)$ that transforms according to the above one-dimensional representation. Explicitly, we consider polynomials of the form
\begin{equation}
    f^{(l)}(\mathbf{k}) = \sum_{i+j+k=N_l} c_{ijk}\,k_x^i k_y^j k_z^k,
    \label{eq:wave_poly}
\end{equation}
with integer exponents $i,j,k\in\mathbb{Z}_{\ge0}$ and total degree $i+j+k=l$. For a given SSG, we determine the smallest integer $l>0$ for which there exists a nontrivial set of coefficients $\{c_{ijk}\}$ such that $f^{(l)}(\mathbf{k})$ transforms as
\begin{equation}
\begin{aligned}
    f^{(l)}(\mathbf{k}) &= +\,f^{(l)}(R^T\mathbf{k}), && \text{for all spin-preserving operations },\\
    f^{(l)}(\mathbf{k}) &= -\,f^{(l)}(R^T\mathbf{k}), && \text{for all spin-flipping operations }.
\end{aligned}
\end{equation}
In group-theoretical terms, this is equivalent to asking whether the representation of the SSG on the space of degree-$l$ polynomials contains a one-dimensional subrepresentation with characters $\pm1$ as specified above. Practically, this reduces to solving a linear function for the coefficients $\{c_{ijk}\}$, and is therefore straightforward to implement numerically. The resulting $l$ agrees with the "characteristic spin-group integer" defined in terms of spin-degenerate nodal surfaces crossing the $\Gamma$ point in Ref.~\cite{smejkal2021altermagnetism}.

We apply this procedure to the bulk coplanar odd-wave and even-wave classes discussed in the main text. For coplanar odd-wave SSGs, we find
\begin{itemize}
    \item $p$-wave: 17,049 SSGs (leading with $l=1$),
    \item $f$-wave: 637 SSGs (leading with $l=3$),
    \item $h$-wave: 116 SSGs (leading with $l=5$),
    \item $j$-wave: 14 SSGs (leading with $l=7$),
    \item $l$-wave: 14 SSGs (leading with $l=9$),
\end{itemize}
while for coplanar even-wave SSGs we find
\begin{itemize}
    \item $d$-wave: 939 SSGs (leading with $l=2$),
    \item $g$-wave: 198 SSGs (leading with $l=4$),
    \item $i$-wave: 36 SSGs (leading with $l=6$).
\end{itemize}

\section{candidates of odd-wave magnet in MAGNDATA database}
\label{sm:p-wave magnet candidates}
In this section, we list candidate odd-wave magnets identified in the MAGNDATA database~\cite{Gallego2016magndataI}. We find 73 entries (including \ch{CeNiAsO}, previously proposed in Ref.\cite{hellenes2024pwavemagnets}) whose SSGs and magnetic configurations realize coplanar odd-wave magnetism of the $p$- and $f$-wave types; higher-order $h$-, $j$-, and $l$-waves, are not found in currently known materials. We also note that the material search and the definition of these phases have been independently proposed in Ref.\cite{oddSSG}.
\begin{table}[h!]
    \centering
    \caption{List of candidate materials, their SSG numbers\cite{jiang2023enumeration}, MAGNDATA numbers, and corresponding even-wave symmetry.}
    \begin{tabular}{llll|llll}
        \hline
        Material & SSG number & MAGNDATA  & Wave 
                 & Material & SSG number & MAGNDATA  & Wave \\
        \hline
        \ch{Ba3MnNb2O9}   & 164.3.2.2.P & 1.0.8   & $f$-wave 
                          & \ch{CsFeCl3}        & 194.3.2.2.P   & 1.0.14 & $f$-wave \\
        \ch{EuIn2As2}     & 194.3.4.3.P & 1.0.32  & $p$-wave 
                          & \ch{CsMnBr3}        & 194.3.4.6.P   & 1.0.35 & $f$-wave \\
        \ch{RbFeCl3}      & 194.3.2.2.P & 1.0.40  & $f$-wave 
                          & \ch{RbNiCl3}        & 194.3.4.6.P   & 1.0.41 & $f$-wave \\
        \ch{CsNiCl3}      & 194.3.4.6.P & 1.0.42  & $f$-wave 
                          & \ch{Ba3CoSb2O9}     & 194.3.4.6.P   & 1.0.44 & $f$-wave \\
        \ch{Ba3CoSb2O9}   & 194.3.4.6.P & 1.0.45  & $f$-wave 
                          & \ch{Ba3Nb2NiO9}     & 164.6.2.3.P   & 1.13   & $f$-wave \\
        \ch{PrMn2O5}      & 26.2.2.7.P  & 1.19    & $p$-wave 
                          & \ch{HoAuGe}         & 8.2.1.6.P     & 1.34   & $p$-wave \\
        \ch{NdNiO3}       & 26.2.2.32.P & 1.44    & $f$-wave 
                          & \ch{HoNiO3}         & 4.2.1.2.P     & 1.48   & $p$-wave \\
        \ch{Cs2CoCl4}     & 14.2.2.4.P  & 1.51    & $p$-wave 
                          & \ch{GdMn2O5}        & 26.2.2.7.P    & 1.54   & $p$-wave \\
        \ch{TmPtIn}       & 25.2.2.27.P & 1.67    & $p$-wave 
                          & \ch{NaNdFeWO6}      & 1.2.1.2.P     & 1.68   & $p$-wave \\
        \ch{BiMn2O5}      & 26.2.2.30.P & 1.74    & $p$-wave 
                          & \ch{DyMn2O5}        & 26.2.2.7.P    & 1.76   & $p$-wave \\
        \ch{DyFe4Ge2}     & 25.2.2.2.P  & 1.98    & $p$-wave 
                          & \ch{TbMn2O5}        & 6.2.1.8.P     & 1.108  & $p$-wave \\
        \ch{HoMn2O5}      & 6.2.1.8.P   & 1.109   & $p$-wave 
                          & \ch{YBaFe4O7}       & 4.2.2.2.P     & 1.124  & $p$-wave \\
        \ch{TmPdIn}       & 174.2.3.1.P & 1.163   & $f$-wave 
                          & \ch{Tm5Ni2In4}      & 6.2.1.8.P     & 1.17   & $p$-wave \\
        \ch{GeCu2O4}      & 119.4.2.3.P & 1.185   & $f$-wave 
                          & \ch{CoNb2O6}        & 18.2.2.5.P    & 1.224  & $p$-wave \\
        \ch{Ca2Cr2O5}     & 46.2.1.6.P  & 1.227   & $p$-wave 
                          & \ch{NiPS3}          & 5.2.1.4.P     & 1.231  & $p$-wave \\
        \ch{VCl2}         & 164.6.2.3.P & 1.237   & $f$-wave 
                          & \ch{VBr2}           & 164.6.2.3.P   & 1.238  & $f$-wave \\
        \ch{SmFe3(BO3)4}  & 155.2.1.2.P & 1.266   & $f$-wave 
                          & \ch{CeNiAsO}        & 11.2.2.2.P    & 1.272  & $p$-wave \\
        \ch{Yb2Cu2O5}     & 7.2.2.5.P   & 1.28    & $p$-wave 
                          & \ch{DyMn2O5}        & 26.2.2.7.P    & 1.324  & $p$-wave \\
        \ch{Yb2CoMnO6}    & 4.2.1.2.P   & 1.328   & $p$-wave 
                          & \ch{Lu2CoMnO6}      & 4.2.1.2.P     & 1.33   & $p$-wave \\
        \ch{LuMnO3}       & 31.2.2.11.P & 1.34    & $f$-wave 
                          & \ch{Sr2FeO3Cl}      & 129.2.4.5.P   & 1.38   & $f$-wave \\
        \ch{Sr2FeO3Br}    & 129.2.4.5.P & 1.381   & $f$-wave 
                          & \ch{Ca2FeO3Cl}      & 129.2.4.5.P   & 1.382  & $f$-wave \\
        \ch{Ca2FeO3Br}    & 129.2.4.5.P & 1.383   & $f$-wave 
                          & \ch{Sr2FeO3F}       & 129.2.4.5.P   & 1.385  & $f$-wave \\
        \ch{Sr2FeO3F}     & 129.2.4.2.P & 1.386   & $f$-wave 
                          & \ch{Sr2FeO3F}       & 129.2.4.5.P   & 1.387  & $f$-wave \\
        \ch{Cu4O3}        & 119.4.2.3.P & 1.418   & $f$-wave 
                          & \ch{Gd2BaCuO5}      & 26.2.2.7.P    & 1.443  & $p$-wave \\
        \ch{Er2Pt}        & 31.2.2.7.P  & 1.444   & $p$-wave 
                          & \ch{Sr2CuO2Cu2S2}   & 139.4.2.5.P   & 1.456  & $f$-wave \\
        \ch{Li2MnGeO4}    & 7.2.2.8.P   & 1.484   & $p$-wave 
                          & \ch{CsFe(MoO4)2}    & 147.6.2.4.P   & 1.499  & $p$-wave \\
        \ch{DyBe13}       & 140.2.2.32.P & 1.517  & $p$-wave 
                          & \ch{TbBe13}         & 140.2.2.32.P  & 1.518  & $p$-wave \\
        \ch{InMnO3}       & 157.2.6.1.P & 1.524   & $f$-wave 
                          & \ch{InMnO3}         & 157.2.6.1.P   & 1.525  & $f$-wave \\
        \ch{TbC2}         & 47.2.2.8.P  & 1.533   & $p$-wave 
                          & \ch{GeNi2O4}        & 166.2.2.2.P   & 1.562  & $p$-wave \\
        \ch{GeCo2O4}      & 166.2.2.2.P & 1.564   & $p$-wave 
                          & \ch{DyMn2O5}        & 6.2.1.2.P     & 1.599  & $p$-wave \\
        \ch{LuNiO3}       & 4.2.1.2.P   & 1.657   & $p$-wave 
                          & \ch{DyGa3}          & 8.2.1.2.P     & 1.658  & $p$-wave \\
        \ch{Ba3CoNb2O9}   & 164.6.2.3.P & 1.665   & $f$-wave 
                          & \ch{KFe(PO3F)2}     & 147.12.2.5.P  & 1.669  & $p$-wave \\
        \ch{NiCr2O4}      & 24.2.2.2.P  & 1.688   & $p$-wave 
                          & \ch{CsCrF4}         & 6.2.1.8.P     & 1.709  & $p$-wave \\
        \ch{Ba3NiTa2O9}   & 164.6.2.3.P & 1.725   & $f$-wave 
                          & \ch{RuCl3}          & 1.2.1.2.P     & 1.726  & $p$-wave \\
        \ch{Cu2MnSiS4}    & 7.2.2.5.P   & 1.73    & $p$-wave 
                          & \ch{Cu2FeSiS4}      & 7.2.2.8.P     & 1.731  & $p$-wave \\
        \ch{Cu2MnGeS4}    & 7.2.2.5.P   & 1.733   & $p$-wave 
                          & \ch{ErAuIn}         & 189.2.6.1.P   & 1.747  & $f$-wave \\
        \ch{TbAuIn}       & 189.2.6.1.P & 1.748   & $f$-wave 
                          &                     &               &        &         \\
        \hline
    \end{tabular}
\end{table}

\nocite{*}

\end{document}